\documentclass[showpacs,aps,prd,reprint,superscriptaddress,longbibliography]{revtex4-1}
\usepackage[colorlinks=true, pdfstartview=FitV, linkcolor=magenta,citecolor=blue, urlcolor=blue,
bookmarks=true, bookmarksnumbered=true, breaklinks]{hyperref}
\usepackage[dvipdfmx]{graphicx}
\usepackage{amsmath,amssymb,amsthm,bm,mathrsfs,tabularx}
\allowdisplaybreaks[3]

\begin{document}
%\title{Roper-like $\Lambda_c(2765)$ and $\Xi_c(2970)$ as a pentaquark state}
\title{Novel pentaquark picture for singly heavy baryons from chiral symmetry}

\author{Daiki~Suenaga}
\email[]{suenaga@rcnp.osaka-u.ac.jp}
\affiliation{Research Center for Nuclear Physics,
Osaka University, Ibaraki, 567-0048, Japan }

\author{Atsushi~Hosaka}
\email[]{{hosaka@rcnp.osaka-u.ac.jp}}
\affiliation{Research Center for Nuclear Physics,
Osaka University, Ibaraki, 567-0048, Japan }
\affiliation{Advanced Science Research Center, Japan Atomic Energy Agency (JAEA), Tokai 319-1195, Japan}

\date{\today}

\begin{abstract}
We propose a new type of structure for singly heavy baryons of $Qqq\bar{q}q$ in addition to the conventional one of $Qqq$. Based on chiral symmetry of the light quarks, we show that the $Qqq\bar{q}q$ baryon offers a novel picture for heavy quark spin-singlet and flavor-antisymmetric baryons. By making use of the effective Lagrangian approach, we find $\Lambda_c(2765)$ and $\Xi_c(2967)$ are mostly $Qqq\bar{q}q$ while $\Lambda_c(2286)$ and $\Xi_c(2470)$ are mostly $Qqq$. The masses of negative-parity baryons are predicted. 
We also derive a sum rule and the extended Goldberger-Treiman relation that the masses of the baryons satisfy. Furthermore, a mass degeneracy of parity partners of the baryons with the restoration of chiral symmetry is discussed. These are the unique features that the conventional picture of radial excitation in the quark model does not accommodate. Our findings provide useful information not only for future experiments but also for future lattice simulations on diquarks.

\end{abstract}


\maketitle

%%%%%%%%%%%%%%%%%%%%%%%%%%%
\section{Introduction}
\label{sec:Introduction}
%%%%%%%%%%%%%%%%%%%%%%%%%%%
Investigation of heavy baryons has been attracting great attention with recent development of experimental observation at, e.g., KEK, LHC, and SLAC. Singly heavy baryons contain only one heavy quark, the mass of which is larger than the typical energy scale of Quantum Chromodynamics (QCD). There the heavy quark behaves not only as a color source but also as a spectator for the remaining light quarks governed by nonperturbative QCD~\cite{Manohar:2000dt}. Therefore, singly heavy baryons provide useful testing ground toward elucidation of colorful light-quark dynamics inside hadrons.

In this regard, understanding of hadrons from chiral symmetry is important, since it is one of the fundamental symmetries of QCD. In fact, the dynamics of light pseudo-scalar mesons and nucleons are systematically formulated by the spontaneous breakdown of chiral symmetry. In addition, the expected mass degeneracy of chiral partners of light hadrons at extreme conditions provides a key signal of the restoration of chiral symmetry for experiments and lattice simulations~\cite{Hatsuda:1994pi,Harada:2003jx}. In this paper we discuss singly heavy baryons from the chiral symmetry point of view which cannot be easily implemented in the quark model~\cite{Copley:1979wj,Yoshida:2015tia}. In particular, we investigate unique features that are derived from the linear representations of chiral symmetry.

For baryons of light flavors, the {\it mirror representation} has been proposed for the negative parity nucleon $N^*(1535)$, referred to as the {\it mirror nucleon}.
This picture contrasts with the quark model description 
of orbital excitation for $N^*(1535)$.  
In the chiral representation approach, not only spectroscopies 
but also the properties at the extreme conditions such as change in masses and a degeneracy reflecting the chiral partner structurecan be studied~\cite{Detar:1988kn,Jido:2001nt,Gallas:2009qp,Yamazaki:2018stk}.
The latter has been strongly suggested by the lattice simulations~\cite{DeTar:1987ar,Aarts:2017rrl}. The quark content of the mirror nucleon is understood as an exotic pentaquark state ($qqq\bar{q}q$).

%%%%%%%%%%%%%%%%%%%%%%%%%%%
\begin{figure}[htbp]
\centering
\includegraphics*[scale=0.22]{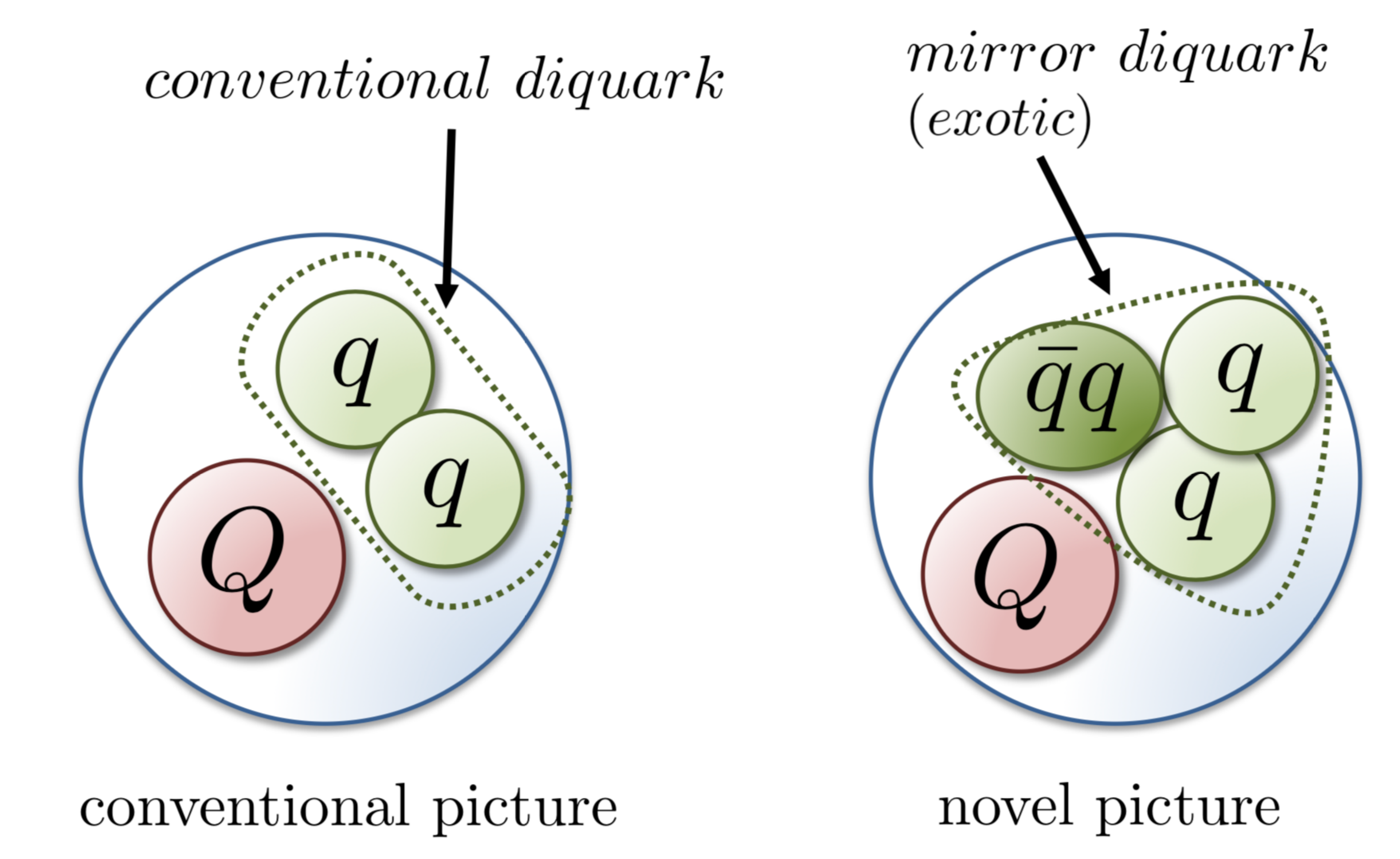}
\caption{A sketch of conventional picture (left) and novel picture proposed in this paper (right) for describing singly heavy baryons. }
\label{fig:Diquark}
\end{figure}
%%%%%%%%%%%%%%%%%%%%%%%%%%%

Turning to the heavy hadrons, the linear representations of chiral symmetry have been studied in various contexts~\cite{Nowak:1992um,Bardeen:1993ae,Dmitrasinovic:2012zz,Ma:2017nik,Kawakami:2019hpp,Dmitrasinovic:2020wye}. Recently $\Xi_c(2967)$ was studied experimentally and its spin and parity were determined to be $J^P=\frac{1}{2}^+$~\cite{Belle:2020tom}. 
The properties of $\Lambda_c(2765)$ are similar to $\Xi_c(2967)$, so that the spin and parity of $\Lambda_c(2765)$ are likely to be $J^P=\frac{1}{2}^+$~\cite{Arifi:2020yfp}. We can safely assume that both the ground state $\Lambda_c(2286)$ [$\Xi_c(2470)$] and the excited state $\Lambda_c(2765)$ [$\Xi_c(2967)$] are heavy quark spin-singlet and flavor-antisymmetric, since no spin partner baryons have been observed~\cite{Zyla:2020zbs}, i.e., the total angular momentum of the light quarks in these heavy baryons is zero: $j_l=0$. Recently observed $\Lambda_b(6072)$ baryon also seems to share the same properties~\cite{Aaij:2020rkw}. 
Within the conventional quark model, $\Lambda_c(2765)$ [$\Xi_c(2967)$] can be understood as a radial excitation. 
In this paper, as an analogue of the mirror nucleon for $N^*(1535)$ we propose a new description where these heavy baryons are 
described by the mirror representation of chiral symmetry.  
As anticipated from the previous paragraph, the mirror representation for heavy baryons is realized by exotic multi-quark contents.  Due to the presence of one heavy quark, it is achieved by the 
{\it mirror diquark}, which is actually a tetraquark $d' \sim qqq \bar q$ as displayed in the right panel in Fig.~\ref{fig:Diquark}. 
The significance of the mirror diquark is implied when we implement the chiral representation for diquarks. Introduction of the mirror diquark enables us to understand $\Lambda_c(2765)$ [$\Xi_c(2967)$] as well as $\Lambda_c(2286)$ [$\Xi_c(2470)$] in a unified way based on chiral symmetry.

The novel state shown in the right panel of Fig.~\ref{fig:Diquark} is distinct from the hadronic molecule state~\cite{Chen:2016qju,Guo:2017jvc}, but should be regarded as a compact pentaquark one because $d'$ is a colorful object binding the remaining heavy quark strongly. 

In the quark model~\cite{Copley:1979wj,Yoshida:2015tia}, $\Lambda_c(2765)$ [$\Xi_c(2967)$] is considered as analogous states of the Roper resonance $N(1440)$~\cite{Roper:1964zza} which have been studied as a radial excitation~\cite{Burkert:2017djo}.\footnote{In the quark model approach, the properties of this Roper-like baryons $\Lambda_c(2765)$ [$\Xi_c(2967)$] such as its mass and decay width can be reproduced by employing the $\,^3P_0$ model~\cite{Chen:2016iyi}, or by including relativistic corrections~\cite{Arifi:2021orx}.} 
In contrast, in our approach by introducing the mirror diquark with chiral symmetry we will find various unique features that cannot be seen in the quark model. Namely $\Lambda_c(2765)$ [$\Xi_c(2967)$] and $\Lambda_c(2286)$ [$\Xi_c(2470)$] are described as superpositions of the three-quark and pentaquark states. Moreover general relations that the masses of the baryons satisfy such as the sum rule and the extended Goldberger-Treiman relation are derived. Furthermore, our approach gives a natural explanation for the strong suppression of direct decay process of $\Lambda_c(2765)$ by the two pion emission~\cite{Arifi:2020yfp}. In addition to those findings, our present approach can provide useful information on hadron properties such as a mass degeneracy of the parity partners~\cite{Suenaga:2014sga,Sasaki:2014asa,Suenaga:2017deu,Buchheim:2018kss,Ishikawa:2019dcn,Montana:2020lfi} at extreme conditions, e.g., finite temperature and/or density where chiral symmetry is (partially) restored.

This paper is organized as follows. In Sec.~\ref{sec:Model} we present an effective Lagrangian for the heavy baryons formed by the conventional and mirror diquarks. In Sec.~\ref{sec:Results} we determine the parameters of our model and show the resultant mass spectrum of the baryons. In Sec.~\ref{sec:Discussion} we discuss the mass degeneracy of the parity partners with the restoration of chiral symmetry. In Sec.~\ref{sec:Conclusions} we conclude our present work.

%%%%%%%%%%%%%%%%%%%%%%%%%%%
\section{Model}
\label{sec:Model}
%%%%%%%%%%%%%%%%%%%%%%%%%%%
%Here, we construct an effective model of singly heavy baryons made of the conventional and mirror diquarks. 
The interpolating fields of the conventional and mirror diquarks contributing to $\Lambda_c(2286)$ [$\Xi_c(2470)$] and $\Lambda_c(2765)$ [$\Xi_c(2967)$] are
\begin{eqnarray}
(d_R)^a_i &\sim& \epsilon_{ijk}\epsilon^{abc}(q_R^T)^b_jC(q_R)^c_k \ , \nonumber\\
(d_L)^a_i &\sim& \epsilon_{ijk}\epsilon^{abc}(q_L^T)^b_jC(q_L)^c_k \ , \nonumber\\
(d'_R)_i^a &\sim&\epsilon_{jkl}\epsilon^{abc}(q_R^T)^b_kC(q_R)^c_l [(\bar{q}_L)^d_i(q_R)^d_j] \ , \nonumber\\
(d'_L)_i^a &\sim&  \epsilon_{jkl}\epsilon^{abc}(q_L^T)^b_kC(q_L)^c_l[(\bar{q}_R)^d_i(q_L)^d_j] \ ,  \label{Diquark}
\end{eqnarray}
where the right- (left-)handed light quark $q_{R(L)}$ is defined by $q_{R(L)} = \frac{1\pm\gamma_5}{2}q$ with $q=(u,d,s)^T$. The subscript ``$i,j, \cdots$'' and the superscript ``$a,b,\cdots$'' indicate the chiral and color indices, respectively. We have introduced scalar diquarks whose total angular momentum is zero: $j_l=0$, here because $\Lambda_c(2286)$ [$\Xi_c(2470)$] and $\Lambda_c(2765)$ [$\Xi_c(2967)$] are heavy quark spin-singlet. Equation~(\ref{Diquark}) shows that $d_R$, $d_L$, $d_R'$, and $d_L'$ belong to the chiral representation of
\begin{align}
d_R\sim  ({\bm 1}, \bar{\bm 3}) \ , \ \  d_L \sim (\bar{\bm 3}, {\bm 1})\ , \nonumber\\
d'_R\sim  (\bar{\bm 3}, {\bm 1}) \ , \ \  d'_L \sim ({\bm 1}, \bar{\bm 3})\ .
\end{align}
Therefore, singly heavy baryons formed by a heavy quark $Q$ and a diquark in Eq.~(\ref{Diquark}) transform as
\begin{align}
B_R \to  B_Rg^\dagger_R \ , \ \  B_L \to B_L g_L^\dagger \ , \nonumber\\
B'_R \to  B'_Rg^\dagger_L \ , \ \  B'_L \to B'_L g_R^\dagger \ ,
\end{align}
with $g_{R(L)} \in SU(3)_{R(L)}$, where $B_R$, $B_L$, $B'_R$, and $B_L'$ are given by
\begin{align}
B_{R,i} \sim Q^a (d_R)_i^a \ , \ \  B_{L,i} \sim Q^a (d_L)_i^a \ , \nonumber\\
B'_{R,i} \sim Q^a (d'_R)_i^a \ , \ \  B'_{L,i} \sim Q^a (d'_L)_i^a \ , \label{Baryons}
\end{align}
respectively. The interpolating fields in Eq.~(\ref{Baryons}) imply that the heavy quark only plays a role of a spectator for light-quark degrees of freedom.

Heavy baryon fields in Eq.~(\ref{Baryons}) allow us to construct an $SU(3)_L\times SU(3)_R$ symmetric effective Lagrangian within the heavy baryon effective theory (HBET) as
\begin{eqnarray}
{\cal L}_{\rm eff} &=&\sum_{\chi=L,R}\Big( \bar{B}_\chi iv\cdot\partial {B}_\chi-\mu_1\bar{B}_\chi B_\chi  \nonumber\\
&+&  \bar{B}'_\chi iv\cdot\partial B'_\chi -\mu_2\bar{B}'_\chi B'_\chi\Big)  \nonumber\\
&-& \mu_3(\bar{B}_{R} B'_{L} + \bar{B}'_{L}B_{R} + \bar{B}_{L} B'_{R} + \bar{B}'_{R}{B}_{L}) \nonumber\\
&-& g_1(\bar{B}_L\Sigma^* B_R+\bar{B}_R\Sigma^TB_L)  \nonumber\\
&-& g_2( \bar{B}'_R \Sigma^* B'_L + \bar{B}'_L\Sigma^TB_R ) \nonumber\\
&-& g_3(\bar{B}'_R\Sigma^*B_R +  \bar{B}_L\Sigma^* B'_L + {\rm h.c.})\ . \label{BLagrangian}
\end{eqnarray}
Here $\Sigma=S+iP$ consists of scalar ($S$) and pseudo-scalar ($P$) meson nonets and transforms as $\Sigma \to g_L\Sigma g_R^\dagger$. Also, $v^\mu$ is a velocity of the heavy baryons. In Lagrangian~(\ref{BLagrangian}), $\mu_1$, $\mu_2$, and $\mu_3$ terms are responsible for the part of heavy baryon mass which are of order $\Lambda_{\rm QCD}$, since a common mass parameter $m_B$ for all the heavy baryons is subtracted for defining the HBET. The remaining $g_1$, $g_2$, and $g_3$ terms are for the interactions between light quarks and light mesons. These terms also contribute to the masses of heavy baryons when chiral symmetry is spontaneously broken. The terms proportional to $\mu_3$, $g_1$ and $g_2$ in Eq.~(\ref{BLagrangian}) break the $U(1)_A$ axial symmetry allowed by the quantum anomaly. We note that the mass term of $\bar{B}^{(\prime)}_RB_R^{(\prime)}+\bar{B}^{(\prime)}_LB_L^{(\prime)}$ is allowed in Eq.~(\ref{BLagrangian}) because the property of spinor for the baryons is determined by the heavy quark which is irrelevant to the chiral representation of the diquarks. The physical baryons as parity eigenstates of $\pm$ are given by 
\begin{eqnarray}
B_\pm &=& \frac{1}{\sqrt{2}}(B_R\mp B_L)\ , \nonumber\\
B_\pm' &=& \frac{1}{\sqrt{2}}(B_L'\mp B_R')\ ,
\end{eqnarray}
with $B^{(\prime)}_R  \leftrightarrow -B^{(\prime)}_L$ under parity transformation~\cite{Harada:2019udr}.

Under the spontaneous breakdown of chiral symmetry, $\Sigma$ acquirers its vacuum expectation values (VEVs) as $\langle \Sigma \rangle = {\rm diag}(\sigma_q,\sigma_q,\sigma_s)$. When we ignore the $u$ and $d$ quark masses, $\sigma_q$ becomes identical to the pion decay constant. We will use $\sigma_q = 93$ MeV as one of inputs for model parameters, whereas $\sigma_s$ is not determined by the decay constants due to corrections from the non-negligible $s$ quark mass $m_s\sim 100$ MeV. By substituting the parity eigenstates $B_\pm$ and $B_\pm'$ into the Lagrangian~(\ref{BLagrangian}), and by reading off the mass terms together with the VEVs $\langle \Sigma\rangle$, mass eigenvalues of the heavy baryons are obtained as
\begin{eqnarray}
 M(B_{+,i}^{H/L}) &=& m_B + \frac{1}{2}\Big[m_{+,i}+m_{+,i}' \nonumber\\
&& \pm\sqrt{(m_{+,i}-m_{+,i}')^2 + 4\tilde{m}^2_{+,i}}\ \Big]\ , \nonumber\\
 M(B_{-,i}^{H/L}) &=& m_B + \frac{1}{2}\Big[m_{-,i}+m_{-,i}' \nonumber\\
&& \pm\sqrt{(m_{-,i}-m_{-,i}')^2 + 4\tilde{m}^2_{-,i}}\ \Big] \ ,\label{LambdaPartnerMass}
\end{eqnarray}
with 
\begin{eqnarray}
m_{\pm,i} &=& \mu_1\mp g_1\sigma_i \ , \nonumber\\
m_{\pm,i}' &=& \mu_2\mp g_2\sigma_i\ , \nonumber\\
\tilde{m}_{\pm,i} &=& \mu_3\mp g_3\sigma_i\ .
\end{eqnarray}
Here $i=1,2,3$ stands for the flavor index where $i=1,2$ corresponds to $\Xi_c$'s while $i=3$ to $\Lambda_c$'s. In Eq.~(\ref{LambdaPartnerMass}) we have introduced a common heavy mass parameter $m_B$ to define the HBET~\cite{Manohar:2000dt}. The choice of $m_B$ is arbitrary, thus we take the average mass of all the heavy baryons which will be explicitly given latter. For the VEVs, we have defined $ \sigma_{i=1}=\sigma_{i=2}\equiv \sigma_q$ and $ \sigma_{i=3}\equiv \sigma_s$. The superscript $H$ $(L)$ represents the higher (lower) mass eigenstate, i.e. $H$ ($L$) corresponds to ``$+$'' (``$-$'') sign in front of the square root in right-hand-side. In obtaining Eq.~(\ref{LambdaPartnerMass}), we have defined mass eigenstates by
\begin{align}
\left(
\begin{array}{c}
B_{\pm,i}^L \\
B_{\pm,i}^H \\
\end{array}
\right) = \left(
\begin{array}{cc}
\cos\theta_{B_{\pm,i}} & \sin\theta_{B_{\pm,i}} \\
-\sin\theta_{B_{\pm,i}} & \cos\theta_{B_{\pm,i}} \\
\end{array}
\right)\left(
\begin{array}{c}
B_{\pm,i} \\
B_{\pm,i}' \\
\end{array}
\right)\ , \label{Mixing}
\end{align}
with the mixing angle satisfying $\tan\theta_{B_{\pm,i}} =(2\tilde{m}_{\pm,i})/(m_{\pm,i}-m'_{\pm,i}) $. Hence the physical states are superpositions of three-quark ($B_{\pm,i}$) and pentaquark ($B_{\pm, i}'$) states as depicted in Fig.~\ref{fig:Diquark}. This is the novel picture of singly heavy baryons proposed in this paper.

%%%%%%%%%%%%%%%%%%%%%%%%%%%
\section{Mass spectrum}
\label{sec:Results}
%%%%%%%%%%%%%%%%%%%%%%%%%%%
To start with, we explain how the remaining parameters $\mu_1$, $\mu_2$, $\mu_3$, $g_1$, $g_2$, $g_3$, and $\sigma_s$ are fixed. First, we employ masses of experimentally observed states of $J^P=\frac{1}{2}^+$; $\Lambda_c(2286)$, $\Lambda_c(2765)$, $\Xi_c(2470)$, and $\Xi_c(2967)$ are used as inputs for~\cite{Zyla:2020zbs}
\begin{eqnarray}
&&M(B_{+,i=3}^L) = 2286\,  {\rm MeV}\  , \ M(B_{+,i=3}^H) = 2765\, {\rm MeV}\ , \nonumber\\
&&M(B_{+,i=1,2}^L) = 2470\, {\rm MeV}\ ,\ M(B_{+,i=1,2}^H) = 2967\, {\rm MeV}\ . \nonumber\\
\end{eqnarray} 
Next, we use the masses of the conventional diquarks which are considered to be  decoupled from the mirror ones. These masses were estimated by the lattice QCD~\cite{Bi:2015ifa} as 
\begin{eqnarray}
&& M(d_{+,i=1,2}) = 906\, {\rm MeV} \ ,\  M(d_{-,i=1,2})=1142\, {\rm MeV}\ , \nonumber\\
&& M(d_{+,i=3}) = 725\, {\rm MeV}\ ,  \ M(d_{-,i=3}) = 1265\, {\rm MeV}\ . \label{DQuarkInput}
\end{eqnarray}
It should be noted that the last one in Eq.~(\ref{DQuarkInput}) is estimated by a chiral effective theory~\cite{Harada:2019udr} together with the simulation~\cite{Bi:2015ifa}. Here ``$\pm$'' and ``$i$'' stand for the parity and flavor indices of the diquark, respectively. These diquark masses are then related to the masses of baryons of three-quark states $B_{\pm,i}$ with the pentaquark states switched off: 
$B'_{\pm,i} = 0$ in the Lagrangian~(\ref{BLagrangian}), leading to $M(B_{\pm,i}) = m_B +m_{\pm,i}$. The mass difference of the heavy baryons may equal that of diquarks, i.e.
\begin{align}
 M(B_{-,i})-M(B_{+,i}) = 2g_1\sigma_i = M(d_{-,i}) - M(d_{+,i})\ .
\end{align}
which can be used to fix $g_1$ and $\sigma_s$.

In addition to the above inputs, we employ a quark model prediction as a useful reference for the demonstration. Namely, we take the mass of the lightest heavy quark spin-singlet $\Lambda_c$ baryon carrying $J^P=\frac{1}{2}^-$ predicted in Ref.~\cite{Yoshida:2015tia} as another input: $M(B_{-,i=3}^L)=2890$ MeV. We note that the observed $J^P=\frac{1}{2}^-$ baryons of $\Lambda_c(2595)$ and $\Xi_c(2790)$ are heavy quark spin-doublet which are not treated in this paper.

%%%%%%%%%%%%%%%%%%%%%%%%%%%
\begin{figure}[htbp]
\centering
\includegraphics*[scale=0.36]{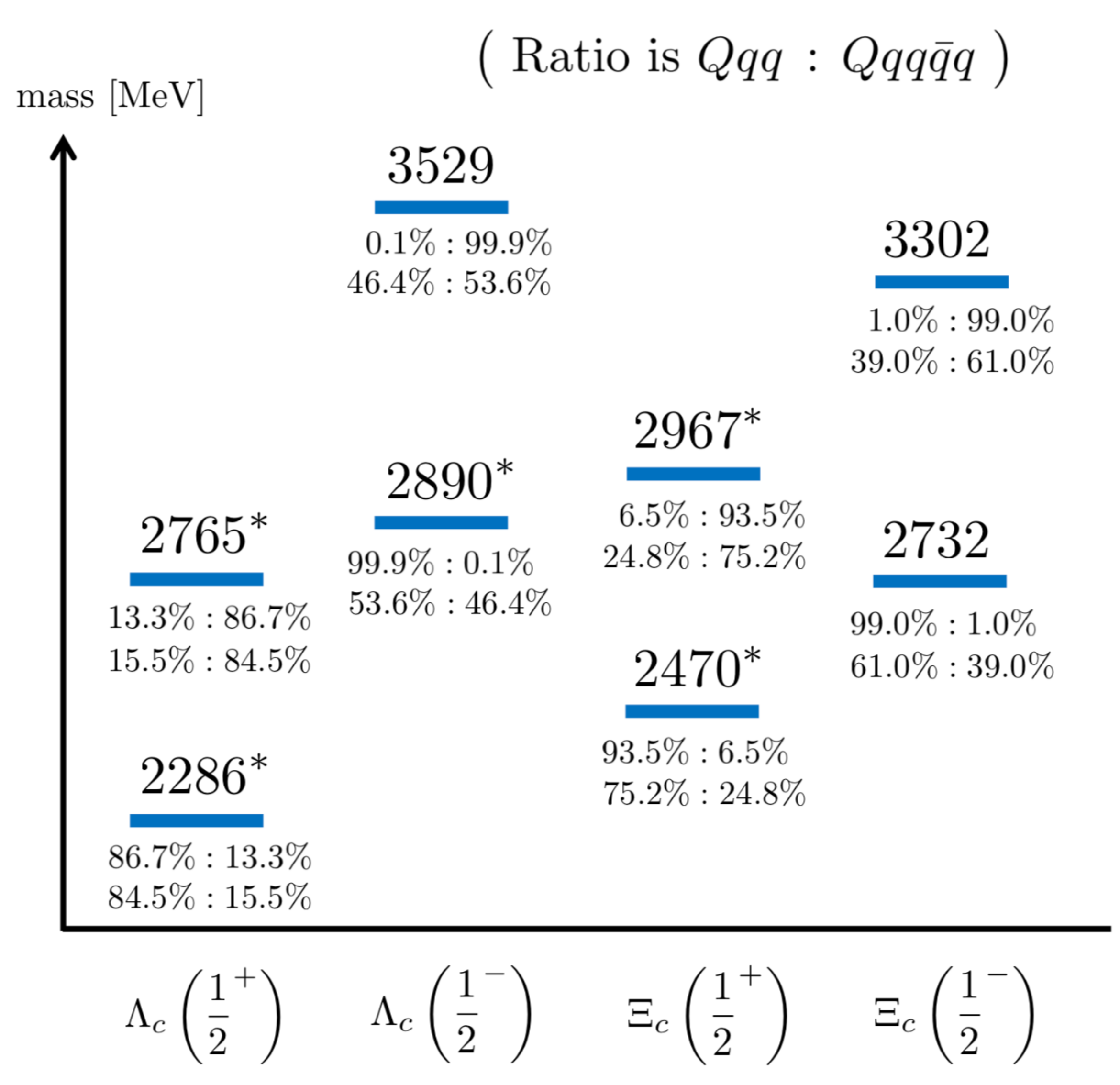}
\caption{Mass spectrum of $\Lambda_c$'s and $\Xi_c$'s with $J^P=\frac{1}{2}^+,\frac{1}{2}^-$ obtained in our model. The asterisk ($*$) stands for the inputs. The ratios indicated under the bars correspond to the components of three-quark and pentaquark states in each baryon: $Qqq : Qqq\bar{q}q$, in which the upper and lower ratios correspond to the parameter sets (I) and (II), respectively.}
\label{fig:CharmedBaryons}
\end{figure}
%%%%%%%%%%%%%%%%%%%%%%%%%%%

%%%%%%%%%%%%%%
\begin{table*}[htbp]
\begin{center}
\renewcommand{\arraystretch}{1.3}
 \begin{tabular}{c|cccccccc} \hline
 & $\mu_1$ [MeV] & $\mu_2$ [MeV] & $\mu_3$ [MeV] & $g_1$ & $g_2$ & $g_3$ & $\sigma_q$ [MeV] & $\sigma_s$ [MeV] \\ \hline
 Set (I) & $-247$ & $247$ & $\mp91.0$ & $1.27$ & 1.94 & $\pm0.34$ & $93^*$ & 212   \\ 
Set (II) & $94.1$ & $-94.1$ & $\pm246$ & $1.27$ & 1.94 & $\pm0.34$ & $93^*$ & 212    \\ \hline
 \end{tabular}
\caption{Two parameter sets (I) and (II). The mass parameter $m_B$ is fixed to be $m_B=2868$ MeV for both the sets. The asterisk ($\ast$) stands for the inputs. }
\label{tab:Parameters}
\end{center}
\end{table*}
%%%%%%%%%%%%%%

%%%%%%%%%%%%%%
\begin{table}[htbp]
\begin{center}
\renewcommand{\arraystretch}{1.3}
 \begin{tabular}{c|cccc} \hline
 & $\theta_{B_{+,i=1,2}}$ & $\theta_{B_{-,i=1,2}}$ & $\theta_{B_{+,i=3}}$ & $\theta_{B_{-,i=3}}$ \\ \hline
 Set (I) & $\pm14.8^\circ$ & $\pm6.01^\circ$ & $\pm21.4^\circ$ & $\pm1.67^\circ$  \\ 
Set (II) & $\pm29.9^\circ$ & $\pm38.6^\circ$ & $\pm23.2^\circ$ & $\pm43.0^\circ$  \\ \hline
 \end{tabular}
\caption{The fixed mixing angles for parameter sets (I) and (II), with the sign corresponding to the ones in the text. Note that $\theta_{B_{\pm,i=3}}$ and $\theta_{B_{\pm,i=1,2}}$ stand for the mixing angles of $\Lambda_c$ baryons and those of $\Xi_c$ baryons, respectively. }
\label{tab:Angles}
\end{center}
\end{table}
%%%%%%%%%%%%%%

Now we can fix all the parameters. Since the mass eigenvalues in Eq.~(\ref{LambdaPartnerMass}) include square roots, the coupled equations yield four solutions for the parameter sets. Physically these four sets are classified into two sets (I) and (II) as displayed in Table~\ref{tab:Parameters}. Both the parameter sets (I) and (II) predict the remaining negative-parity baryon masses as $M(B_{-,i=1,2}^{L})=2732$ MeV, $M(B_{-,i=1,2}^{H})=3302$ MeV, and $M(B_{-,i=3}^{H})=3529$ MeV. We show the resultant mass spectrum of the heavy baryons in Fig.~\ref{fig:CharmedBaryons} where the asterisk ($*$) stands for the inputs. Also, the mixing angles defined in Eq.~(\ref{Mixing}) are determined as shown in Table~\ref{tab:Angles}, where $\theta_{B_{\pm,i=3}}$ and $\theta_{B_{\pm,i=1,2}}$ stand for the mixing angles of $\Lambda_c$'s and those of $\Xi_c$'s, respectively. Table~\ref{tab:Angles} indicates that, for $J^P=\frac{1}{2}^+$ baryons the lower state $\Lambda_c(2286)$ [$\Xi_c(2470)$] is dominated by $Qqq$ while the higher one $\Lambda_c(2765)$ [$\Xi_c(2967)$] by $Qqq\bar{q}q$ for both the parameter sets. On the other hand, for $J^P=\frac{1}{2}^-$ baryons the ratio is largely dependent on the parameters. The ratio of the $Qqq$ and $Qqq\bar{q}q$ components for each baryon is also shown under the bars in Fig.~\ref{fig:CharmedBaryons}, in which the upper and lower ratios correspond to the parameter sets (I) and (II), respectively. The ratio is estimated by the square of each coefficient in Eq.~(\ref{Mixing}), i.e. $\cos^2\theta_{B_{\pm,i}}$ or $\sin^2\theta_{B_{\pm,i}}$.

   %%%%%%%%%%%%%%%%%%%%%%%%%%%%%%%%
\begin{figure*}[t]
  \begin{center}
    \begin{tabular}{cc}

      % 1
      \begin{minipage}[c]{0.5\hsize}
       \centering
       \hspace*{-2.5cm} 
         \includegraphics*[scale=0.68]{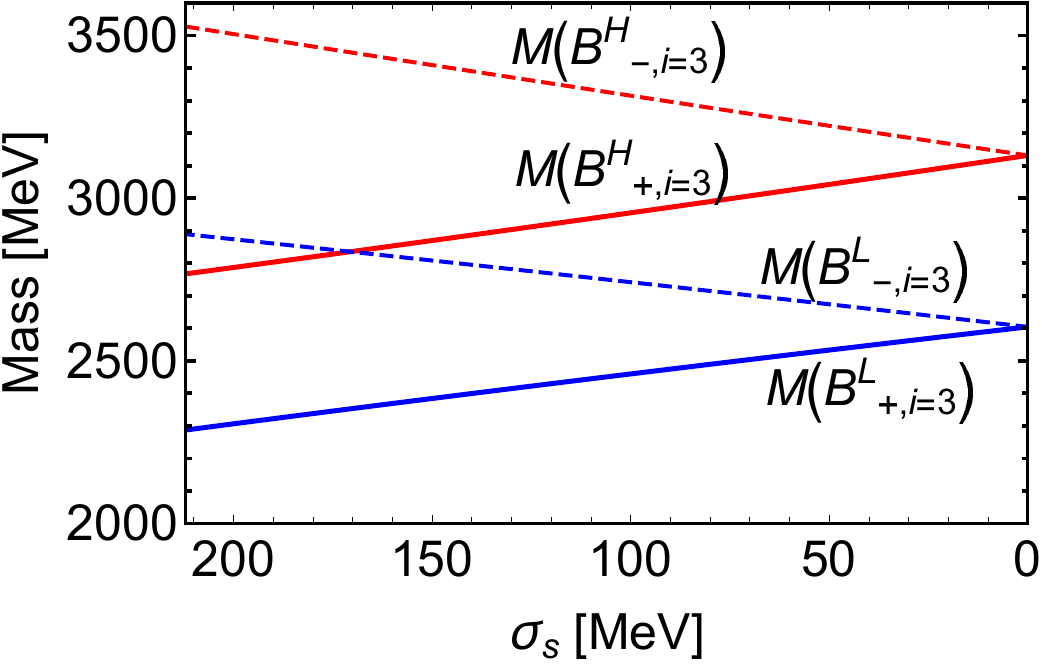}\\
%          \hspace{-1cm} (a) Spectral function for pion $\rho_\pi^*(q_0,\vec{q})$
         \end{minipage}

      % 2
      \begin{minipage}[c]{0.4\hsize}
       \centering
        \hspace*{-0.3cm} 
          \includegraphics*[scale=0.68]{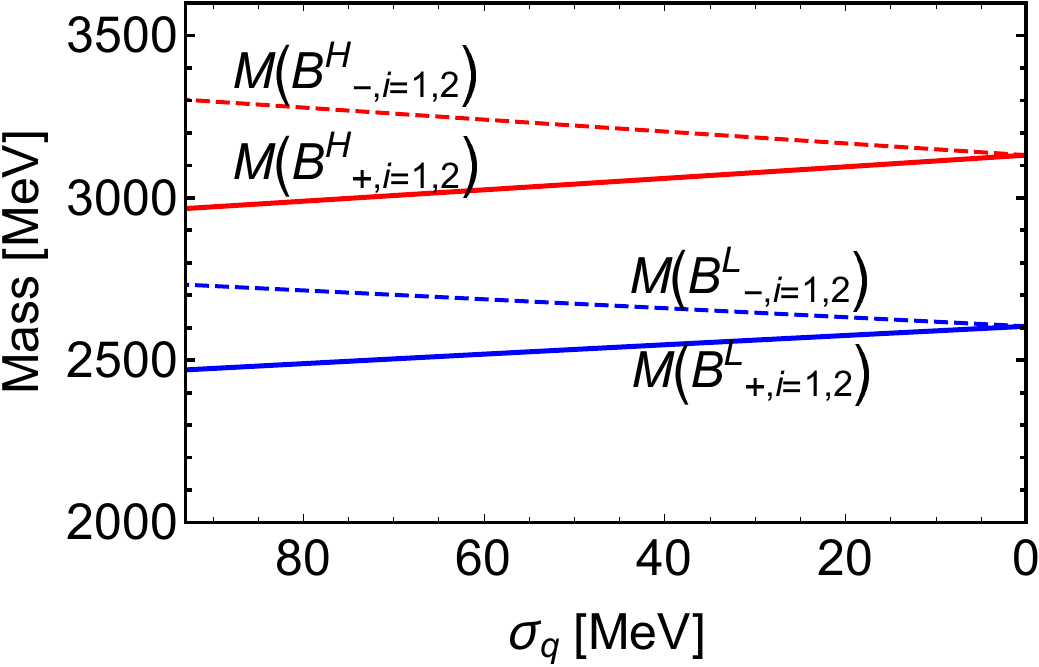}\\
%          \hspace{-1cm} (b) Spectral function for $\sigma$ meson $\rho_\sigma^*(q_0,\vec{q})$
      \end{minipage}

      \end{tabular}
 \caption{The left panel shows the $\sigma_s$ dependence of $M(B_{+,i=3}^L)$, $M(B_{-,i=3}^L)$, $M(B_{+,i=3}^H)$, and $M(B_{-,i=3}^H)$, while the right one shows the $\sigma_q$ dependence of $M(B_{+,i=1,2}^L)$, $M(B_{-,i=1,2}^L)$, $M(B_{+,i=1,2}^H)$, and $M(B_{-,i=1,2}^H)$. The results are identical for the parameter sets (I) and (II) in Table~\ref{tab:Parameters}.} 
\label{fig:MassMod1}
  \end{center}
\end{figure*}
%%%%%%%%%%%%%%%%

The $\Lambda_c(2765)$ [$\Xi_c(2967)$] baryon have been found to be mainly $Qqq\bar{q}q$ with $J^P=\frac{1}{2}^+$ where the $\bar{q}q$ constituent in the mirror diquark $qq\bar{q}q$ requires a $P$-wave excitation. Although adding such a $P$-wave $\bar{q} q$ pair costs an energy of order 1 GeV in the quark model, the above analysis shows that the mirror diquark based on chiral symmetry costs only 
about 0.5 GeV to form $\Lambda_c(2765)$ [$\Xi_c(2967)$]. This finding is similar to the small mass of the light scalar meson $\bar{q}q$ in the chiral model.

One of the most important consequences of our model is the sum rule of the masses:
\begin{align}
\sum_{p=\pm,n=H,L} M(B_{p,i=1,2}^n) = \sum_{p=\pm,n=H,L} M(B_{p,i=3}^n)\ , \label{SumRule}
\end{align}
which can be derived from Eq.~(\ref{LambdaPartnerMass}). Namely the sum of mass of the four $\Lambda_c$'s coincides with that of the four $\Xi_c$'s. The mass formula~(\ref{LambdaPartnerMass}) also yields the extended Goldberger-Treiman relation
\begin{eqnarray}
\sum_{n=H,L}M(B_{-,i}^n)-\sum_{n=H,L}M(B_{+,i}^n)= 2(g_1+g_2)\sigma_i \nonumber\\
\label{GTRelation} 
\end{eqnarray}
for each $i$, which gives a constraint on the mass difference between parity partners and the coupling constant of one pion (kaon) emission. In deriving Eq.~(\ref{GTRelation}) the higher and lower masses $M^H_{p,i}$ and $M^L_{p,i}$ have been summed up to cancel out the square roots in Eq.~(\ref{LambdaPartnerMass}).

At the end of this section we give comments on decay properties of the excited baryons in our present model. In this paper, we have investigated mostly the mass spectrum of the baryons as the first step towards understanding of the importance of the newly introduced pentaquark from chiral symmetry. In addition to the masses, our Lagrangian~(\ref{BLagrangian}) can derive various couplings among the light mesons and heavy baryons. Namely, decay properties of the excited baryons can be also studied. In this case, the ground-state $\Sigma_c$ baryons are necessary since the excited baryons can decay into them. Based on chiral symmetry, the $\Sigma_c$ baryons are straightforwardly included on top of our present model~\cite{Harada:2019udr}.\footnote{We expect that inclusion of a pentaquark component for the ground-state $\Sigma_c$ baryons is not necessary since they are not excited states.} By comparing our calculation given by such a hybrid model and the experimental data of decays, e.g., the $\Lambda_c(2765)$ decay widths, some of the model parameters are expected to be fixed or constrained such that the uncertainty of our present model is narrowed. Moreover, to shed light on the decay properties of the excited baryons predicted in this paper is useful for future heavy baryon experiments to observe them. In addition, detailed understanding of decay properties of the excited baryons, especially from the viewpoint of the Goldberger-Treiman relation~(\ref{GTRelation}) and $U(1)_A$ anomaly~\cite{Kawakami:2020sxd}, leads to further elucidation of properties of exotic constituents from symmetry aspects of QCD. Investigation of the decays is beyond the scope of this paper and we leave it for future publication.

%For instance, the decay channel of $\Lambda_c(2765)$ is $\Lambda_c(2765)\to\pi\pi$ as denoted in PDG. Here, the previous work by an angular correlation analysis of the decay~\cite{Arifi:2020yfp} shows that the experimental data of the decay can be explained by only sequential process of $\Lambda_c(2765)\to \Sigma_c(2455)\pi$ and $\Lambda_c(2765)\to \Sigma_c(2520)\pi$.

%%%%%%%%%%%%%%%%%%%%%%%%%%%
\section{Baryon masses with the restoration of chiral symmetry}
\label{sec:Discussion}
%%%%%%%%%%%%%%%%%%%%%%%%%%%
The symmetry relations such as the sum rule~(\ref{SumRule}) and the extended Goldberger-Treiman relation~(\ref{GTRelation}) provide useful information on the chiral symmetry properties of the baryons for future experiments and lattice simulations. In addition, as one of the most important advantages of employing the present chiral effective model, we can examine the properties of baryons at extreme conditions such as finite temperature/density where chiral symmetry tends to be restored. The mirror diquark can be regarded as an analogue of the mirror nucleon~\cite{Detar:1988kn,Jido:2001nt,Gallas:2009qp,Yamazaki:2018stk}, which gives rise to the mass degeneracy of parity partners at the chiral restoration point in the nucleon sector~\cite{DeTar:1987ar,Zschiesche:2006zj,Motohiro:2015taa,Aarts:2017rrl,Suenaga:2017wbb,Ishikawa:2018yey}. Thus, we can expect that a similar mass degeneracy of the partners of the singly heavy baryons arises.

%In this section, in order to clarify the unique prediction of our present model we discuss a qualitative behavior of the mass modifications of the baryons at such an  environment. 
In order to see the above mass degeneracy, we study mass modifications of the baryons by changing the VEVs $\sigma_q$ and $\sigma_s$ in the mass formulae~(\ref{LambdaPartnerMass}). In Fig.~\ref{fig:MassMod1} we show the VEV dependence of the baryon masses. The left panel of this figure shows the $\sigma_s$ dependence of $M(B_{+,i=3}^L)$, $M(B_{-,i=3}^L)$, $M(B_{+,i=3}^H)$, and $M(B_{-,i=3}^H)$, while the right one shows the $\sigma_q$ dependence of $M(B_{+,i=1,2}^L)$, $M(B_{-,i=1,2}^L)$, $M(B_{+,i=1,2}^H)$, and $M(B_{-,i=1,2}^H)$. It should be noted that the results are identical for the parameter sets (I) and (II) in Table~\ref{tab:Parameters}. Figure~\ref{fig:MassMod1} shows that 
\begin{eqnarray}
&& M(B_{+,i}^L) = M(B_{-,}^L) \ , \nonumber\\
&& M(B_{+,i}^H) = M(B_{-,i}^H) \label{Partners}
\end{eqnarray}
hold for $i=1,2$ and $i=3$, respectively, at the chiral restoration point denoted by $\sigma_q=0$ and $\sigma_s=0$. Therefore, our model predicts that $B_{+,i}^L$ and $B_{-,i}^L$ are the parity partners as in the nucleon sector. Similarly, $B_{+,i}^H$ and $B_{-,i}^H$ are the partners. Even when chiral symmetry is restored, due to the presence of $\mu_3$ which is chiral invariant, still the baryons are interpreted as superpositions of a three-quark state and a pentaquark state. The role of the mass parameter $\mu_3$ is similar to the one of the so-called {\it chiral invariant mass} $M_0$ for the naive and mirror nucleons~\cite{Detar:1988kn,Jido:2001nt}. Namely, $\mu_3$ and $M_0$ are expected to share common fundamental properties.

The mass degeneracy demonstrated above originating from the restoration of chiral symmetry is expected to be realized at extreme conditions such as finite temperature/density. These environments are provided in heavy-ion collisions (HICs) and lattice simulations. Further studies on the properties of the baryons at such extreme conditions from the viewpoint of the restoration of chiral symmetry are left for future work.

%%%%%%%%%%%%%%%%%%%%%%%%%%%
\section{Conclusions}
\label{sec:Conclusions}
%%%%%%%%%%%%%%%%%%%%%%%%%%%
In this paper, we have proposed a new type of the light-quark degrees of freedom named mirror diquark in addition to the conventional diquark, to explain $\Lambda_c(2765)$ [$\Xi_c(2970)$] and $\Lambda_c(2286)$ [$\Xi_c(2470)$] in a unified way based on chiral symmetry. Accordingly, we have obtained the masses of negative-parity as well as positive-parity baryons in Eq.~(\ref{LambdaPartnerMass}), and furthermore have derived the unique relations such as the sum rule~(\ref{SumRule}) and the extended Goldberger-Treiman relation~(\ref{GTRelation}). Moreover our model can naturally explain the strong suppression of the direct decay process of $\Lambda_c(2765)$ by the two pion emission~\cite{Arifi:2020yfp}, because the $\sigma$-$\Lambda_c(2765)$-$\Lambda_c(2286)$ ($\sigma$ is the light scalar meson) coupling disappears after the diagonalization in the Lagrangian~(\ref{BLagrangian}). %\cmb{We also have shown that the mass degeneracy of the parity partners~(\ref{Partners}) takes place at the chiral restoration point, whose precursory behavior is expected to be tastable in future HIC experiments and lattice simulations.}

The mass spectrum of singly heavy baryons predicted in this paper will provide useful information for future experiments including HICs. Moreover, our finding that the mirror diquark plays a significant role is expected to provide new direction for future lattice simulations on diquarks~\cite{Hess:1998sd,Alexandrou:2006cq,Babich:2007ah,Bi:2015ifa}. For example, the importance of the mirror diquark would be tested by examining the correlators of the conventional $qq$ and mirror $qq\bar{q}q$ diquarks, similarly to the simulation of light scalar mesons with $\bar{q}q$ and $\bar{q}q\bar{q}q$ states~\cite{Wakayama:2014gpa}. The development of the diquark chiral effective theory~\cite{Hong:2004xn} including the mirror diquark would be of interest. We also expect that the mirror diquark with chiral symmetry provides a new aspect for the understanding of the Roper resonance $N(1440)$ with the unique relations similar to Eqs.~(\ref{SumRule}) and~(\ref{GTRelation}).

%The mirror diquark can be regarded as an analogue of the mirror nucleon~\cite{Detar:1988kn,Jido:2001nt,Gallas:2009qp,Yamazaki:2018stk}, which gives rise to the mass degeneracy of parity partners in the nucleon sector at extreme conditions where chiral symmetry is restored~\cite{DeTar:1987ar,Zschiesche:2006zj,Motohiro:2015taa,Aarts:2017rrl,Suenaga:2017wbb,Ishikawa:2018yey}. In fact, in the present model we can confirm that the mass degeneracy of parity partners of singly heavy baryons will take place at chiral symmetry restoration point: $M(B_{+,i}^H) = M(B_{-,i}^H)$ and $M(B_{+,i}^L) = M(B_{-,i}^L)$ from Eq.~(\ref{LambdaPartnerMass}) with $\sigma_q=\sigma_s=0$.

In our novel picture, the mirror diquark is the main constituent of $\Lambda_c(2765)$ [$\Xi_c(2970)$], while conventionally such a baryon is regarded as a radial excitation ($2S$ state) in the quark model~\cite{Arifi:2020ezz}. Namely, an additional mixing from $2S$ state to the pentaquark for $\Lambda_c(2765)$ [$\Xi_c(2970)$] is expected~\footnote{In addition, another constituent provided by bi-local interpolating fields~\cite{Dmitrasinovic:2011yf,Chen:2013efa} may enter.}. Therefore, checking the unique relations Eqs.~(\ref{SumRule}) and~(\ref{GTRelation}), and the mass degeneracy of parity partners or its precursory behaviors in future experiments or lattice simulations would be desired for the better understanding of the hadrons from chiral symmetry.

%%%%%%%%%%%%%%%%%%%%%%%%%%%
\section*{Acknowledgement}
%%%%%%%%%%%%%%%%%%%%%%%%%%%
We thank Ahmad Jafar Arifi for fruitful discussions and comments. We also thank Veljko Dmitrasinovic for useful comments. A. H. is supported in part by Grants-in Aid for Scientific Research, Grants
No. 17K05441(C) and by Grants-in Aid for Scientific Research on Innovative Areas (No.
18H05407).

\bibliography{reference}

\end{document}